\documentclass[aps,prl,twocolumn,superscriptaddress,floatfix]{revtex4}
\usepackage{graphicx}
\usepackage{psfrag}
\usepackage{color}
\usepackage{subfigure}
\usepackage{amsmath}
\definecolor{dred}{rgb}{0.7,0.0,0.0}
\allowdisplaybreaks 
 
\begin{document}

%
% Title Page
%

\title{Competing Pairing Symmetries in a Generalized Two-Orbital Model \\ 
for the Pnictide Superconductors}

\author{Andrew Nicholson}
\author{Weihao Ge}
\author{Xiaotian Zhang}
 
\affiliation{Department of Physics and Astronomy, The University of
  Tennessee, Knoxville, TN 37996} 
\affiliation{Materials Science and Technology Division, Oak Ridge
  National Laboratory, Oak Ridge, TN 32831} 

\author{Jos\'e Riera}

\affiliation{Instituto de F\'isica Rosario, Universidad Nacional de Rosario, 2000-Rosario, Argentina}

%Consejo Nacional de Investigaciones
%Cient\'ificas y T\'ecnicas, 

\author{Maria Daghofer}

\affiliation{IFW Dresden, P.O. Box 27 01 16, D-01171 Dresden, Germany}

\author{Andrzej M. Ole\'s}

\affiliation{Max-Planck-Institut fur Festkorperforschung, 
Heisenbergstrasse 1, 70569
Stuttgart, Germany}
\affiliation{M. Smoluchowski Institute of Physics, Jagellonian
University, Reymonta 4, 30-059 Krak\'ow, Poland}

\author{George B. Martins}

\affiliation{Department of Physics, Oakland University, Rochester, 
Michigan 48309}

\author{Adriana Moreo}
\author{Elbio Dagotto}

\affiliation{Department of Physics and Astronomy, The University of
  Tennessee, Knoxville, TN 37996} 
\affiliation{Materials Science and Technology Division, Oak Ridge
  National Laboratory, Oak Ridge, TN 32831}

\date{\today}
%\maketitle

\begin{abstract}
We introduce and study an extended ``$t$-$U$-$J$'' two-orbital model for the
pnictides that includes Heisenberg terms deduced from the
strong coupling expansion. Including these 
$J$ terms explicitly allows us to enhance 
the strength of the $(\pi,0)$-$(0,\pi)$
spin order which favors the presence of tightly bound pairing states even in the
small clusters that are here exactly diagonalized.
The $A_{\rm 1g}$ and $B_{\rm 2g}$ pairing symmetries are found to compete
in the realistic spin-ordered and metallic regime.
The dynamical pairing susceptibility additionally unveils low-lying $B_{\rm 1g}$ 
states, suggesting that small changes in parameters 
may render any of the three channels stable.
%These results
%contribute to understanding the puzzling 
%results in pnictides where both nodeless and nodal states have been
%reported.

% pacs of 2008 PRL: \pacs{74.20.Mn, 71.27.+a, 74.20.Rp, 74.70.-b}

\end{abstract}

\maketitle

%\section {Introduction} 

{\it Introduction.} One of the main puzzles in 
iron-based superconductors~\cite{johnston} arises from the
conflicting experimental results on the presence of
nodes in the superconducting state.
Surface-sensitive angle-resolved 
photoemission (ARPES) studies~\cite{arpes} 
indicate that full nearly momentum-independent gaps open on all Fermi 
surface (FS) pockets.
%, as in the $s\pm$ state~\cite{teo}. 
However, some bulk experiments
give results compatible with nodal superconductivity~\cite{nodal}. 
On the theory side, calculations where many different pairing
states are allowed to compete, as opposed to studying a few isolated states, 
are difficult for multiorbital Hubbard models. 
%Beyond mean-field methods, and 
Within magnetic mechanisms for superconductivity, 
two approaches have
addressed this issue: (i) random phase approximation (RPA) 
studies suggested that several
pairing channels are in competition~\cite{graser}, 
in agreement with 
(ii) two-orbital model Lanczos studies~\cite{Daghofer:2009p1970,moreo} 
based on the quantum numbers of the state with two more electrons 
than half-filling. 
However, RPA relies on a particular subset
of diagrams, while
studying the quantum numbers of clusters 
did work before for the cuprates~\cite{static}, but 
finding true Cooper pair formation is difficult.

In this
Letter, a simple generalization of Hubbard models for
pnictides is presented that increases the strength of the $(\pi,0)$-$(0,\pi)$ spin order
in the undoped limit, creating tightly bound-states upon electronic doping  
that can be studied with Lanczos methods on the small clusters currently accessible 
with state-of-the-art computers. These extra terms can be justified 
by noting that the undoped state, which combines itinerant electrons with a robust
N\'eel temperature and a lattice distortion~\cite{dai}, is itself rather exotic
and suggests a more stable magnetic order than a description based exclusively on onsite
Coulomb repulsion would support.
The $t$-$J$ model for cuprates~\cite{dagottoRMP}
provides further guidance: 
here $J$, when considered as independent of $t$,
can be increased to sufficiently large values that 
$d$-wave pairing tendencies are amplified and tightly 
bound-states are formed, while in the
Hubbard model the $d$-wave pairing signal is weak~\cite{dagottoRMP}. 
In pnictides, the strong coupling $t$-$J_1$-$J_2$ model
has been studied before~\cite{si}.
In the alternative ``$t$-$U$-$J$'' route~\cite{tuj} to be followed here, 
Heisenberg ``$J$'' terms will be added 
to the original Hubbard model to enhance spin order and
pairing tendencies, but without projecting 
out doubly occupied sites and charge fluctuations. 
%This approach allows 
%for the intermediate and large interaction $U$ regions to appear 
%in the same formalism, and supports an interesting competition
%between several pairing channels.

{\it Model and method.} The two-orbital 
model~\cite{Daghofer:2009p1970,moreo,raghu} based on the $d_{xz}$ ($x$) and
$d_{yz}$ ($y$) Fe orbitals is studied here. 
Keeping only these two orbitals is reasonable since
$x$ and $y$  provide the largest contribution to the pnictides' band 
structure FS~\cite{phonon}. In addition, the studies described
below are computationally demanding and they simply {\it cannot} be carried out 
with more orbitals. Thus, a balance must be reached 
between the more ideal five-orbital models and
the feasibility of the actual calculations. 
The model includes a hopping term with amplitudes that 
fit band calculations~\cite{raghu} (energy scale $|t_1|$), 
a Hubbard term with
on-site intraorbital repulsion $U$, a Hund coupling $J_{\rm H}$, an interorbital
repulsion fixed as $U'=U-2J_{\rm H}$, 
and a pair-hopping term with coupling $J'$=$J_{\rm H}$~\cite{oles83}. 
The model 
%does not need to be reproduced here since it 
was used in several previous studies~\cite{moreo}. The novelty
are the extra Heisenberg terms that will be added, with
associated exchange couplings for nearest-neighbor (NN) and next-nearest-neighbor (NNN), $J_{\rm NN}$ and $J_{\rm NNN}$, as discussed below.

The two-orbital Hamiltonian 
is exactly investigated using the Lanczos algorithm 
(including dynamical information) on a small
tilted $\sqrt{8}$$\times$$\sqrt{8}$ cluster~\cite{dagottoRMP,Daghofer:2009p1970,moreo}.
This requires substantial computational resources: 
to determine the undoped-limit ground state of the 8-sites cluster 
even exploiting the Hamiltonian symmetries still 
requires a basis with $\sim$2-20~M states (equivalent to a 16-sites cluster
one-band Hubbard model) depending on the subspace explored. 
Lanczos runs had to be
performed for all allowed momenta ${\bf k}$, quantum numbers under
rotations and reflections 
(i.e. irreducible representations $A_{\rm 1g}$, $A_{\rm 2g}$, $B_{\rm 1g}$, $B_{\rm 2g}$, and $E$ 
of the $D_{\rm 4h}$ group~\cite{moreo,commentref}), 
and $z$-axis total spin projections. In addition, binding energies require calculations for 
a number of electrons $N$ equal to 16, 17, and 18, 
varying $U$, $J_{\rm H}$, $J_{\rm NN}$, 
and $J_{\rm NNN}$ 
in a fine grid. The full effort amounted to $\sim$8,000 diagonalizations of
the cluster, supplemented by dynamical calculations,
using a Penguin 128GB Altus 3600 computer.

\begin{figure}[thbp]
\begin{center}
\includegraphics[width=7.6cm,clip,angle=0]{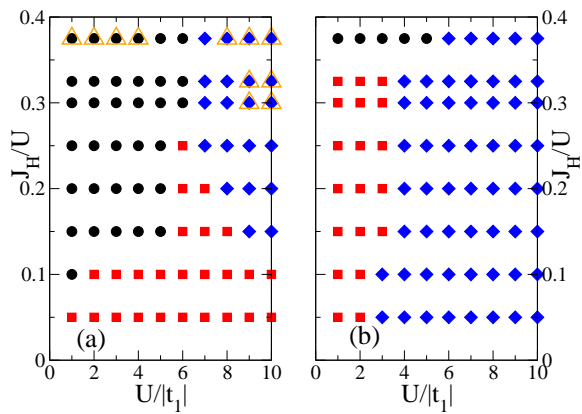}
%\vskip -0.3cm
\caption{(color online)  Relative symmetry between the $N$=16 (undoped) and 
$N$=18 GS's, varying $U$ and $J_{\rm H}/U$.
Circles denote triplet states, squares 
$B_{\rm 2g}$-symmetric singlets, and diamonds $A_{\rm 1g}$-symmetric singlets.
(a) Results for couplings $J_{\rm NN}$=$J_{\rm NNN}=0$. 
Open triangles indicate binding. 
(b) Results for the lowest value of ($J_{\rm NN},J_{\rm NNN}$) 
%at each $(U,J_{\rm H}/U)$, 
where binding appears.}
\vskip -0.5cm
\label{Fig1}
\end{center}
\end{figure}

{\it Results for the original two-orbital model.} 
The relative symmetry between the undoped ($N$=16) ground state (GS) and 
the $N$=18 GS has been studied varying $U$/$|t_1|$ and $J_{\rm H}/U$. The results
at $J_{\rm NN}$=$J_{\rm NNN}$=0 are shown in Fig.~\ref{Fig1}(a). The undoped GS
has momentum ${\bf k}=(0,0)$ and it
transforms according to the $A_{\rm 1g}$ representation of the $D_{\rm 4h}$ group, for
all the investigated values of $J_{\rm H}$ and $U$. The $N$=18 GS also has
${\bf k}=(0,0)$, but its irreducible representation varies in 
different regions of the phase diagram.
In agreement with previous results~\cite{moreo}, the $N$=18 GS
is a spin triplet for $U$$\leq$~6~$|t_1|$\cite{t1} 
and a broad range of $J_{\rm H}/U$.
A spin-singlet state with
symmetry $B_{\rm 2g}$ dominates the small $J_{\rm H}/U$ (roughly $\leq$0.15) 
region for the studied values of $U$/$|t_1|$. 
This $B_{\rm 2g}$ state arises from the orbital portion of the pairing operator since the 
two extra electrons added to the undoped GS 
are located in different orbitals~\cite{moreo}. 
At $U$$\geq$~7~$|t_1|$ and intermediate to 
large $J_{\rm H}/U$ regimes, the singlet-state symmetry becomes $A_{\rm 1g}$. 
The binding energy 
$E_{\rm B}$=$E(18)+E(16)-2E(17)$, where $E(N)$ is the GS
energy of $N$
electrons, was also studied. 
Binding, i.e. $E_{\rm B}<0$,  is observed 
%(triangles in Fig.~\ref{Fig1}(a))
but only at large $U$'s where the undoped GS is an insulator. 

{\it Binding stabilization.} 
As discussed in the Introduction, 
here the spin background with wavevectors $(\pi,0)$-$(0,\pi)$ 
will be magnified via the addition of extra Heisenberg terms
with the expectation that carrier attraction will become stronger, leading
to $E_{\rm B}<0$ pairing. To find the precise form of the Heisenberg terms, 
a strong coupling (large $U$ and $J_{\rm H}$) expansion of the undoped two-orbital Hamiltonian was carried out. NN and NNN Heisenberg interactions $J_{\rm NN}$ and
$J_{\rm NNN}$ were obtained. 
This strong coupling expansion for multiorbital models is subtle and 
it has to be performed in such a way 
that the results are independent of the basis chosen for the orbitals. 
Following, e.g., \cite{oles}
these additional interactions are
$\sum_{\langle{\bf i}{\bf j}\rangle,\alpha \beta}J_{ij}{\bf S}_{{\bf i}\alpha}
\cdot{\bf S}_{{\bf j}\beta},$
%\begin{equation}
%\sum_{<{\bf i},{\bf j}>,\alpha,\beta}J_{i,j}{\bf S}_{{\bf i},\alpha}
%\cdot{\bf S}_{{\bf j},\beta},
%\label{4}
%\end{equation}
%\noindent 
where $\langle{\bf i}{\bf j}\rangle$ indicates NN and NNN sites with 
$J_{ij}$=$J_{\rm NN}$ and $J_{\rm NNN}$, 
$\{\alpha,\beta\}$ label the orbitals ($x$ or $y$), 
and ${\bf S}_{{\bf i}\alpha}$ is the total spin at site ${\bf i}$ and orbital
$\alpha$. The couplings are 
$J_{\rm NN}$=${2(t_1^2+t_2^2)\over{3(U+J_{\rm H})}}$, and 
$J_{\rm NNN}$=${4(t_3^2+t_4^2)\over{3(U+J_{\rm H})}}$, where $t_i$ are the 
hoppings $t_1$=$-1$, $t_2$=$1.3$, 
$t_3$=$t_4$=$-0.85$ in the usual notation~\cite{raghu} 
($|t_1|$ units). $J_{\rm NN}$ and $J_{\rm NNN}$ 
are the {\it same} for both orbitals. 
The ratio ${J_{\rm NN}\over{J_{\rm NNN}}}$=${1\over{2}}{t_1^2+t_2^2\over{t_3^2+t_4^2}}$=$0.93$ is kept fixed 
since $J_{\rm NN}$ and $J_{\rm NNN}$ are both antiferromagnetic, and a ratio $\sim$1 
introduces frustration favoring $(\pi,0)$-$(0,\pi)$ spin order, as 
confirmed by calculating the spin structure factor and varying ${J_{\rm NN}}/{U}$~\cite{suple}.
Below, $J_{\rm NN}$ will be considered a {\it free} parameter independent of 
$U$ and $J_{\rm H}$ (with ${J_{\rm NN}/{J_{\rm NNN}}}$=$0.93$), 
to further enhance such a spin order. 

\begin{figure}[thbp]
\begin{center}
\vskip -0.65cm
\includegraphics[width=8.8cm,clip,angle=0]{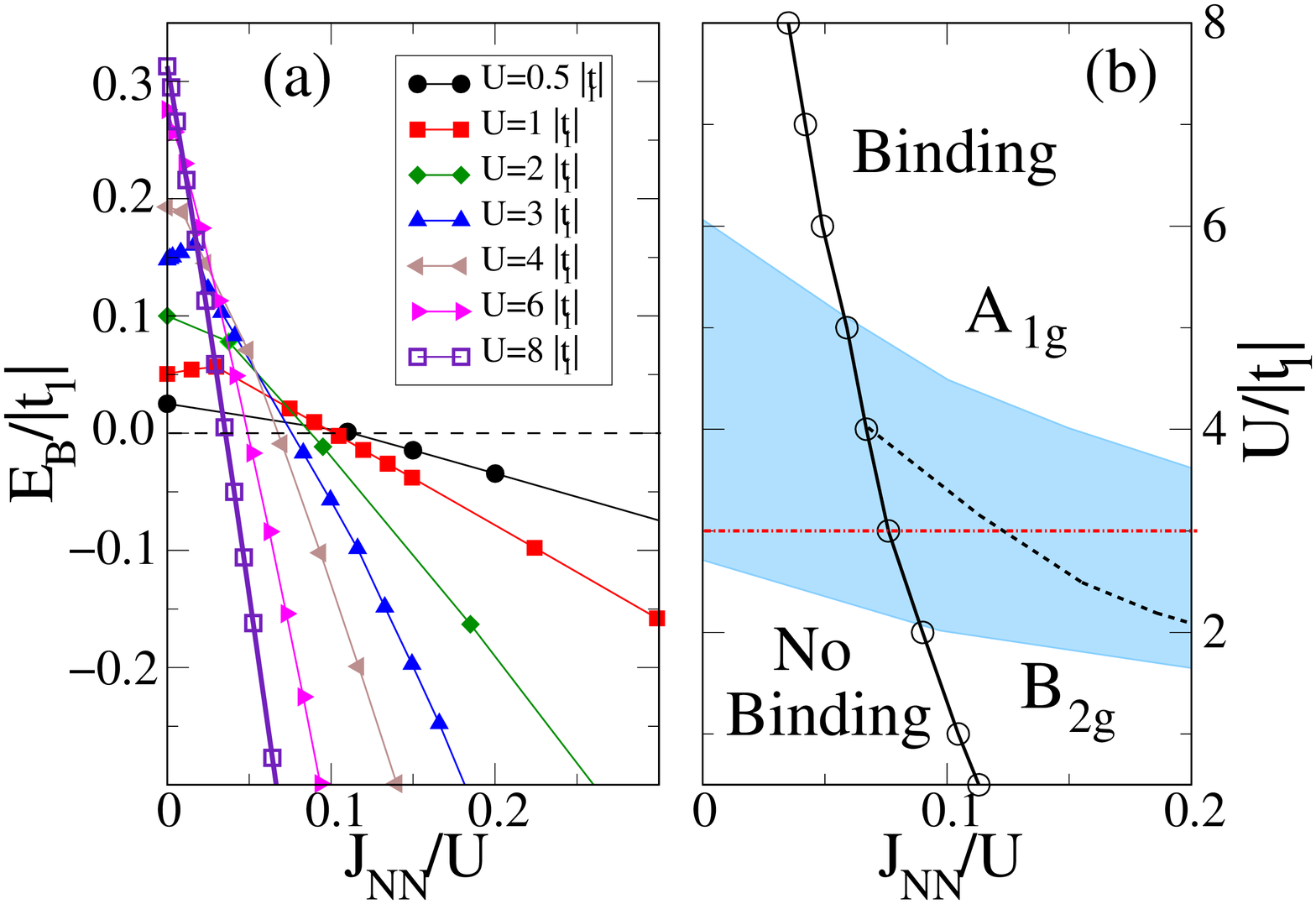}
\vskip -0.3cm
\caption{(color online) (a) $E_{\rm B}$/$|t_1|$ {\it vs.} 
$J_{\rm NN}/U$ for different 
values of $U$/$|t_1|$ and $J_{\rm H}/U$=$0.2$. 
(b) Phase diagram showing ``Binding'' and ``No Binding'' 
regions and the symmetry of the two-electron 
bound state varying
$U$/$|t_1|$ and $J_{\rm NN}/U$, for $J_{\rm H}/U$=$0.2$. 
The shaded area is where the
antiferromagnetic/metallic state is stabilized in the mean-field approximation
for the undoped limit. 
The dot-dashed line is for 
 Fig.~\ref{Fig3}.}
\vskip -0.6cm
\label{Fig2}
\end{center}
\end{figure}

By adding the extra Heisenberg terms to the two-orbital model, 
the desired goal is obtained since 
with increasing $J_{\rm NN}$ eventually $E_{\rm B}$ becomes negative 
for all the studied $(U,J_{\rm H})$ couplings. 
The spin-triplet region virtually disappears (Fig.~\ref{Fig1}(b))
%and it is confined to small/intermediate values of $U$ and large values of $J/U$ 
%(circles in Fig.~\ref{Fig1}(b)).
%The spin-triplet region was 
and it is
mainly replaced by the $B_{\rm 2g}$ state which itself becomes 
confined to $U<4$~$|t_1|$ (squares) due to the expansion of the $A_{\rm 1g}$ 
region.
% expands and reaches now down to 
%intermediate $U\approx 4$ for the whole range of $J/U$. 
%This indicates that
%when terms that induce binding are added the electron doped pairing state with
%symmetry A$_{1g}$ becomes stable for values of $U$ and $J$ that may capture 
%the physics of some pnictides.
%
%At $J_{\rm NN}$=$J_{\rm NNN}$=$0$,
%the $N$=$18$ GS has several very low-lying energy 
%excited states with different symmetries at each $(U,J_{\rm H})$ point. 
%These states tend to become the GS sequentially as $J_{\rm NN}$ increases. 
%Thus, 
The symmetries shown in Fig.~\ref{Fig1}(b) were 
obtained with the smallest superexchange values $(J_{\rm NN}^*,J_{\rm NNN}^*)$ 
that produce binding of 
two electrons at each $(U,J_{\rm H})$ point. 
%We found that with increasing values of the Heisenberg couplings
%the A$_{1g}$ symmetry with binding of electrons is stabilized in the whole 
%phase diagram except for $U\le 1$.
%For example, for $U=1$ and $J/U=0.2$ we 
%observed that the triplet is replaced by a singlet with B$_{2g}$ symmetry 
%where binding of electrons with  $E_B=-0.0023$ occurs for $J_{NN}=0.1046$ 
%and $J_{NNN}=0.1124$ 
%as it can be seen in Fig.~\ref{Fig1}b.
%Increasing the Heisenberg couplings to $J_{NN}=0.3736$ 
%and $J_{NNN}=0.4014$
%the binding increases with $E_B=-0.2187$ while with even larger values of the
%couplings the symmetry changes to B$_{1g}$.
In Fig.~\ref{Fig2}(a), where the binding energy {\it vs}. $J_{\rm NN}/U$ is shown
at several $U$'s and at a fixed (realistic) $J_{\rm H}/U$=$0.2$, 
some examples of $(J_{\rm NN}^*,J_{\rm NNN}^*)$ can be found. 
Increasing $J_{\rm NN}$ eventually induces binding for 
all $U$'s. The value of $J_{\rm NN}/U$ for which binding 
occurs decreases as $U$ increases.

%\vskip -0.3cm

Studying $E_{\rm B}$ and the relative symmetry between the $N$=16 and 18 GS's,  
phase diagrams in 
the $(U,J_{\rm NN}/U)$ plane were constructed.
%for several values of $J_{\rm H}/U$. 
In Fig.~\ref{Fig2}(b), typical
results for $J_{\rm H}/U$=$0.2$ are shown~\cite{foot1}. 
 %Binding eventually 
%develops for all values of $U$, increasing $J_{\rm NN}/U$. 
The bound state has $A_{\rm 1g}$ symmetry
in most of the binding region, but a $B_{\rm 2g}$ symmetric state also 
prevails at small $U$ values ($\sim$ 2 $|t_1|$). Both symmetries appear inside 
the proper magnetic/metallic region of the undoped limit, according to 
mean-field calculations~\cite{rong} extended to incorporate $J_{\rm NN}$.
In~\cite{suple} it is shown that the
results in Figs.~\ref{Fig2}(a,b) are qualitatively the same 
varying ${J_{\rm NN}}/{J_{\rm NNN}}$ in the range [0.5,1.5].

{\it Overlaps.} Consider now the pairing operators that produce the
electronic bound states. The overlap
$\langle\Psi(N$=$18)|\Delta^{\dagger}_{{\bf k},i}|\Psi(N$=$16)\rangle$
%\over{\sqrt{\langle\Psi(N=16)|\Delta\Delta^{\dagger}|\Psi(N=16)\rangle,}}}
was calculated, where
$|\Psi(N)\rangle$ is the GS in the subspace of $N$ 
electrons and 
$\Delta^{\dagger}_{{\bf k},i}$=$\sum_{\alpha \beta} f({\bf k})(\sigma_i)_{\alpha \beta}
d^{\dagger}_{{\bf k},\alpha,\uparrow}d^{\dagger}_{{\bf k},\beta,\downarrow}$,
with $d^{\dagger}_{{\bf k},\alpha,\sigma}$ creating
an electron with spin $z$-axis projection $\sigma$, 
at orbital $\alpha=x,y$, and with momentum ${\bf k}$.
The structure factor $f({\bf k})$ arises from the spatial 
location of the electrons forming the pair~\cite{moreo}, and $\sigma_i$ are the 
Pauli matrices ($i=1,2,3$) or the $2\times 2$ identity matrix $\sigma_0$ 
($i=0$) (note that $\sigma_1$ and $\sigma_2$ imply an interorbital pairing). 
Overlaps for
all the symmetries in \cite{moreo}, 
and with NN and NNN locations for the electronic pairs, 
were evaluated.  

\begin{figure}[t!]
\begin{center}
\includegraphics[width=8.0cm,clip,angle=0]{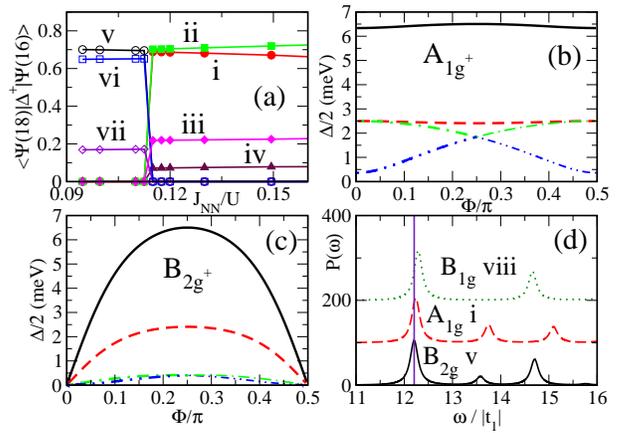}
%\vskip -0.3cm
\caption{(color online) (a) Overlap 
$\langle\Psi(N$=$18)|\Delta^{\dagger}_{{\bf k},i}|\Psi(N$=$16)\rangle$ {\it vs.} 
$J_{\rm NN}/U$ for 
the indicated pairing 
operators, at $U$=$3$~$|t_1|$ and $J_{\rm H}/U$=$0.2$. (b) Superconducting 
gap at the FS: internal hole pocket 
(continuous line), 
external hole pocket (dashed line). 
The dot-dashed and double dot-dashed lines are for the two electron
pockets which intersect at the Brillouin zone boundary ( $\Phi$=$\pi/4$) of the folded 
zone. The $A_{\rm 1g^+}$ symmetric linear combination 
of $A_{\rm 1g}$ operators {\rm (i)} 
and {\rm (ii)} is used, with equal 
weight. The angle $\Phi$ is measured from the positive 
$x$-axis to the positive 
$y$-axis. (c) Same as (b) but 
for the $B_{\rm 2g^+}$ symmetric combination of the
$B_{\rm 2g}$ operators {\rm (v)} and {\rm (vi)}. (d) Dynamic pairing 
susceptibility for the 
pairing operators indicated (see text), at $U$=$3$~$|t_1|$, 
$J_{\rm H}/U$=$0.2$, and $J_{\rm NN}/U$=$0.095$. The vertical line indicates 
$E(18)-E(16)$.}
\vskip -0.7cm
\label{Fig3}
\end{center}
\end{figure}

For all 
operators respecting the relative symmetry between the doped and undoped 
states, finite overlaps were found, although of different values. 
As a trend, as the binding grows, pairing involving
NNN operators prevail over the NN ones.
For example, in the $A_{\rm 1g}$ region in
Fig.~\ref{Fig2}(b) there are four pairing operators with finite overlap 
(shown in Fig.~\ref{Fig3}(a) for $U$=$3$$|t_1|$ and $J_{\rm H}/U$=$0.2$)
characterized by $f({\bf k})\sigma_i$ equal to: 
{\rm (i)} $(\cos k_x+\cos k_y)\sigma_0$
(full circles); {\rm (ii)} $(\cos k_x\cos k_y)\sigma_0$ (full squares); 
{\rm (iii)} $(\sin k_x\sin k_y)\sigma_1$ (full diamonds); and {\rm (iv)}
$(\cos k_x-\cos k_y)\sigma_3$ (full triangles). 
Close to the boundary with the $B_{\rm 2g}$ phase 
where the binding is weak ($E_{\rm B}$$\approx$-0.05~$|t_1|$), 
operators {\rm (i)} and {\rm (ii)} 
present the largest, and almost equal, overlaps.
With increasing binding 
the {\rm (i)} overlap 
decreases while {\rm (ii)} becomes stronger. 
The overlaps for operators {\rm (iii)} and {\rm (iv)} 
are clearly smaller.

%\begin{figure}[thbp]

Note that {\rm (ii)} is the simplest expression of a 
nodeless $s\pm$ pairing operator~\cite{teo}. 
Our results indicate that this type of 
pairing dominates {\it only} when the binding energy 
is large, which occurs at very large $U$ or $J_{\rm NN}$. 
At intermediate values of couplings,
a {\it symmetric} linear combination of {\rm (i)} 
and {\rm (ii)} with almost equal weights is optimal, and it 
leads to a ``quasi-nodal''  $s\pm$ pairing state
(Fig.~\ref{Fig3}(b)). From this perspective, 
the most ``natural'' $A_{\rm 1g}$ pairing
operator arises from a linear combination of {\rm (i)} 
and {\rm (ii)}, as opposed to just {\rm (ii)} as in $s\pm$ scenarios.
% where a strong momentum dependence of the
%gap is observed on the electron pockets. 
The gaps in Fig.~\ref{Fig3}(b) were 
calculated from mean-field approximations as in \cite{moreo} and \cite{our3}, and choosing a pairing strength $V_0$ such that
the gap order-of-magnitude in meV's agrees with experiments~\cite{t1}. 
Note that the linear combination $A_{\rm 1g^+}$ for the hole pockets 
closely reproduces (full and dashed lines) the ARPES results 
 in the superconducting state, with both gaps only weakly 
${\bf k}$-dependent, and with the interior (exterior) pocket gap $\sim$ 12 (6) 
meV. 
%
%The main difference with ARPES 
%occurs for the electron pockets (dot-dashed and 
%double dot-dashed lines) which present strongly ${\bf k}$-dependent gaps with 
%a quasi-node  (maximum) on the $x$-axis for the external (internal) 
%pocket, after
%folding, and a maximum (quasi-node) on the $y$-axis. 
%
The electron pockets, on the other hand, 
present strongly ${\bf k}$-dependent gaps, and
a quasi-node is found along 
the $x$- ($y$-) axes for the pocket at $X$ ($Y$). In the folded
zone, this implies that the quasi node is on the outer pocket,
in agreement 
with angle-resolved specific heat measurements~\cite{nature}.

Note that the presence of a $d_{xy}$ ``patch'' on the electron pockets has
been discussed before by many groups as possibly responsible
for gap nodes (or minima) on the electron pockets. The present results show
that such a minimum (or nodes) can arise {\it without} such an $xy$-patch,
which is important to assess the impact of the various orbitals.
The one-particle spectral function $A({\bf k},\omega)$ was also
calculated~\cite{suple}. Features on the
scale of the magnetic or superconducting gaps cannot be resolved within the
few momenta available, but
the higher energy features at intermediate couplings
are similar to non-interacting bands~\cite{rong}, 
in agreement with ARPES experiments and with local density plus dynamical mean-field theory. 
calculations~\cite{lda+dmft}.

As mentioned before, 
in physically relevant portions  of the phase diagram~\cite{rong} 
%e.g. $U$$\sim$~2~$|t_1|$ at $J_{\rm NN}$=0,
the pairing symmetry  $B_{\rm 2g}$ competes with $A_{\rm 1g}$. 
Three $B_{\rm 2g}$ 
pairing operators with finite overlaps were found in this region: 
{\rm (v)} $(\cos k_x+\cos k_y)\sigma_1$,
{\rm (vi)} $(\cos k_x\cos k_y)\sigma_1$,
and {\rm (vii)} $(\sin k_x\sin k_y)\sigma_0$. 
From Fig.~\ref{Fig3}(a) the interorbital operators {\rm (v)} 
and {\rm (vi)} 
have a much larger GS overlap than the intraorbital operator {\rm (vii)}. 
The mean-field calculation of the gaps for the symmetric combination of the prevailing
$B_{\rm 2g}$ pairing operators, i.e. {\rm (v)}+{\rm (vi)}, is 
in Fig.~\ref{Fig3}(c). All the gaps have nodes along the $x$ and $y$ axes,
also in good agreement with \cite{nature}.
A strong ${\bf k}$ dependence is observed for all FS pockets, and
the electron-pocket gaps are small ($\sim$ 1 meV).

{\it Dynamical Pair Susceptibilities.} To complete our analysis 
the dynamical pair susceptibilities, defined by
$P(\omega)=\int_{-\infty}^{\infty}dt e^{i\omega t}\langle\Delta_{{\bf k},i}(t)
\Delta^{\dagger}_{{\bf k},i}(0)\rangle$, were also studied in the state with $N$=16 
for the pairing operators $\Delta_{{\bf k},i}$.
A procedure 
used in the context of the cuprates will be followed~\cite{jose}. 
Results for $U$=$3$~$|t_1|$, $J_{\rm H}/U$=$0.2$, and several values of $J_{\rm NN}/U$ were obtained 
along the dot-dashed 
line (red) of Fig.~\ref{Fig2}(b). The overlaps calculation already 
indicated that for $N$=$18$ there are several 
low-lying energy states with different symmetries near the GS. The 
dynamical pair susceptibilities show that most of these low lying states
have a large overlap with $\Delta^{\dagger}_{{\bf k},i}|\Psi_{N=16}(0)\rangle$ for  
$\Delta^{\dagger}_{{\bf k},i}$ with the appropriate symmetry. This is 
{\it qualitatively different} to the cuprates' $t$-$J$ model, where the overlap of the doped GS with
$\Delta^{\dagger}_{{\bf k},i}|\Psi(0)\rangle$ was large for $\Delta$ with $d$-wave symmetry
but negligible for $s$-wave symmetry~\cite{jose}. 
In that $s$-wave case the 
spectral weight in $P(\omega)$ accumulates at high energies, 
while $P(\omega)$ for the $d$-wave pairing operator showed a well 
defined sharp peak at the GS energy of the doped 
state~\cite{jose}. This is not the case for the two-orbital model.
For example, in Fig.~\ref{Fig3}(d) at $J_{\rm NN}/U$=$0.095$, where the
doped GS has symmetry $B_{\rm 2g}$, a sharp peak occurs in $P(\omega)$
for the 
$B_{\rm 2g}$ pairing operator {\rm (v)}, but 
a similar behavior is found 
in $P(\omega)$ for the $A_{\rm 1g}$ pairing operator {\rm (i)} 
(the low-lying peak  originates in
a low-lying excited state with $A_{\rm 1g}$ symmetry). In addition, the susceptibility
for a pairing operator {\rm (viii)} $(\cos k_x+\cos k_y)\sigma_3$, NN version of the
$B_{\rm 1g}$ operator {\rm (ix)} $(\cos k_x\cos k_y)\sigma_3$,
is also competitive~\cite{foot} (Fig.~\ref{Fig3}(d)).

{\it Conclusions.} 
The effects of NN and NNN Heisenberg terms on the symmetry and
 the binding energy of two electrons added to the undoped state of the two-orbital
 Hubbard model were studied using Lanczos techniques on small clusters.
%The symmetry 
%and the binding energy of two electrons added to 
%the undoped state of a generalized ``$t$-$U$-$J$'' 
%two-orbital model were 
%studied using Lanczos techniques on small clusters. 
%We found 
%that in order to obtain a GS with bound electrons for values of the 
%parameters in the physically relevant regime, we need to add NN and
%NNN Heisenberg couplings which arise from a perturbative expansion in the 
%large $U$ limit. We believe that these terms could be dynamically generated 
%in larger systems. 
%Electron binding is eventually achieved in the whole 
%phase diagram ($U$ vs. $J$) as the strength of the added $J_{\rm NN}$ and
%$J_{\rm NNN}$ couplings, and concomitantly the $(\pi,0)$ spin background, increases.
Quasi-nodal $A_{\rm 1g}$ bound states 
are stabilized for physical values 
of $J_{\rm H}/U$, in the intermediate/large $U$ region, in agreement with
RPA results~\cite{graser}. 
%Thus, the 
%assumption of an $A_{\rm 1g}$ pairing symmetry 
%is reasonable as long as the Hubbard
%repulsion is robust.
Our results also
indicate that a competing $B_{\rm 2g}$ state may become stable in  
physically relevant regimes of $U$/$|t_1|$.
% (while the $B_{\rm 1g}$ symmetry that characterizes 
%the cuprates is stable only in unphysical regimes: (i) For the smallest values 
%of $U$ at very large $J_{\rm NN}/U$, and (ii) for very large values of the 
%Heisenberg couplings in all the phase diagram). 
%This $B_{1g}$ state is, 
%nevertheless, not the same as the one 
%observed in the cuprates because the pairing interaction has different 
%sign for each orbital, i.e., the $B_{1g}$ symmetry arises from the orbital 
%rather than the spacial part of the pairing operator.
In addition, the pairing susceptibility presents low-lying
excitations with 
$B_{\rm 2g}$, $A_{\rm 1g}$, and $B_{\rm 1g}$ symmetries.
%\cite{commentRPA}
%,  with large overlap
%with the doped GS for all the values 
%of the parameters studied. 
Thus, pairing correlations with any 
of these symmetries could be stabilized by small modifications
in the model parameters, in 
agreement with ~\cite{graser,Daghofer:2009p1970,moreo,si}.  
This suggests that a similar sensitivity to small details 
may occur among different compounds of the pnictide family.
%, or
%between surface and bulk behaviors, qualitatively different
%from the cuprates where the $d$-wave symmetry is robust.

%\section{Acknowledgments}
 
%{\it Acknowledgments.} 
This work was supported by 
the U.S. DOE, Office of Basic Energy Sciences,
Materials Sciences and Engineering Division (A.N., W.G., X.Z., G.M., A.M., E.D.), 
CONICET, Argentina (J.R.), the DFG under the Emmy-Noether program (M.D.), and by
the Foundation for Polish Science (FNP) and the
Polish government Project N202 069639 (A.M.O.). Conversations with D.-X. Yao and
Thomas Prestel are
acknowledged.


\begin{thebibliography}{10}


\bibitem{johnston} D. C. Johnston, Adv. Phys. {\bf 59}, 803 (2010).
%, and references therein.
%%arXiv:1005.4392.

%\bibitem{pnictides} Y. Kamihara, {\it et al.}, J. Am. Chem. Soc {\bf 130},
%   3296  (2008).
% G.~F. Chen,{\it et al.}, Phys. Rev. Lett.
%  {\bf 101},  057007  (2008); G.~F. Chen,{\it et al.}, Phys. Rev. Lett. 
%{\bf 100},  247002  (2008); H.-H. Wen,{\it et al.}, Europhys. Lett. {\bf 82},
%  17009  (2008); X.~H. Chen,{\it et al.}, Nature {\bf 453},761  (2008);
%Z. Ren, {\it et al.}, Mater. Res. Innovat. {\bf 12},  105  (2008);
%R. Zhi-An, {\it et al.}, Chin. Phys. Lett. {\bf 25}, 2215  (2008);
%Z.-A. Ren, {\it et al.}, Europhys. Lett. {\bf 83},  17002  (2008).

\bibitem{arpes} See for instance 
T. Kondo {\it et al.}, Phys. Rev. Lett. {\bf 101}, 147003 
(2008); H. Ding {\it et al.}, EPL {\bf 83},  47001  (2008). 
%K. Nakayama, {\it et al.},
%Europhys. Lett. {\bf 85},  67002  (2009); L. Wray, {\it et al.},
%arXiv:0808.2185;  D. Hsieh, {\it et al.}, arXiv:0812.2289 (2008).


\bibitem{nodal}
%L. Shan, {\it et al.}, Europhys.  Lett. {\bf 83},  57004  (2008);
%M. Gang, {\it et al.}, Chin. Phys. Lett.{\bf 25},  2221  (2008); 
%C. Ren, {\it et al.}, arXiv:0804.1726, 2008;
%Y. Nakai, {\it et al.},  J. Phys. Soc. Jpn. {\bf 77},  073701  (2008);
%Y.-L. Wang, {\it et al.},  Supercond. Sci. Technol. {\bf 22},  015018  (2009);
%K. Matano, {\it et al.}, Europhys. Lett. {\bf 83},  57001  (2008);H
%X.~L. Wang,{\it et al.}, arXiv:0808.3398, (2008).
%See for instance 
%K. Ahilan {\it et al.}, Phys. Rev. B {\bf 78},  100501  (2008);
H.-J. Grafe {\it et al.}, Phys. Rev. Lett. {\bf 101},  047003  (2008);
J.K. Dong {\it et al.}, Phys. Rev. Lett. {\bf 104}, 087005 (2010).
%K. Hashimoto {\it et al.}, Phys. Rev. B {\bf 82}, 014526 (2010).
%H. Mukuda, {\it et al.}, J. Phys. Soc. Jpn. {\bf 77},  093704  (2008);
%O. Millo {\it et al.}, Phys. Rev. B {\bf 78},  092505  (2008).
%B. Zeng, {\it et al.}, arXiv:1006.2785.

%\bibitem{disorder} V. Mishra, {\it et al.}, arXiv:0901.2653.


%\bibitem{res} P. Szabo, {\it et al.}, Physica {\bf B404}, 3220 (2009) 
%and  Phys. Rev. B {\bf 79}, 012503 (2009); B. Muschler, {\it et al.}, 
%arXiv:0910.0898; V. Stanev, {\it et al.}, Phys. Rev. B {\bf 78},  
%184509 (2008); G. R. Boyd, {\it et al.}, Phys. Rev. B {\bf 79}, 174521 (2009).

\bibitem{graser} S. Graser {\it et al.},
%, T.A. Maier, P.J. Hirschfeld, and D.J. Scalapino,
New J. Phys. {\bf 11}, 025016 (2009).


\bibitem{Daghofer:2009p1970}
M. Daghofer {\it et al.}, Phys. Rev. Lett. {\bf 101},  237004  (2008).

\bibitem{moreo}
A. Moreo {\it et al.}, Phys. Rev. B {\bf 79},
  134502  (2009).



\bibitem{static} E. Dagotto {\it et al.},
%, A. Moreo, F. Ortolani, D. Poilblanc, and J. Riera,
Phys. Rev. B {\bf 45}, 10741 (1992).

\bibitem{dai} C. de la Cruz {\it et al.}, Nature (London) {\bf 453}, 899 (2008).

\bibitem{dagottoRMP} E. Dagotto, Rev. Mod. Phys. {\bf 66}, 763 
(1994).

\bibitem{si} Q. Si and E. Abrahams, Phys. Rev. Lett. {\bf 101}, 076401 (2008);
K. Seo, B.A. Bernevig, and J. Hu, Phys. Rev. Lett. {\bf 101}, 206404 (2008).

\bibitem{tuj} L. Arrachea and D. Zanchi, Phys. Rev. B {\bf 71}, 064519 (2005),
and references therein.

\bibitem{raghu} S. Raghu {\it et al.}, Phys. Rev. B {\bf 77},  220503  (2008).

\bibitem{phonon}
%S. Lebegue, Phys. Rev. B {\bf 75},  035110  (2007);
%D.~J. Singh and M.-H. Du, Phys. Rev. Lett. {\bf 100},  237003  (2008);
%G. Xu, {\it et al.}, Europhys. Lett. {\bf 82},  67002  (2008);
%C. Cao, {\it et al.},  Phys. Rev. B {\bf 77},  220506
%  (2008).WHICH ONE HAS THE ORBITAL COMPOSITION?
L. Boeri, O.V. Dolgov, and A.A. Golubov, 
Phys. Rev. Lett. {\bf 101},  026403 (2008).
%H.-J. Zhang, {\it et al.}, Chin. Phys. Lett. {\bf 26},  017401  (2009). 

\bibitem{oles83} A.M. Ole\'s, Phys. Rev. B \textbf{28}, 327 (1983).

%\bibitem{Kru09} F. Kr\"uger {\it et al.\/}, \prb \textbf{79}, 054504 (2009).

\bibitem{commentref} A similar symmetry analysis can be used to improve cluster dynamical mean-field theory calculations, 
see E. Koch {\it et al.}, Phys. Rev. B {\bf 78}, 115102 (2008).

\bibitem{t1} From comparisons with \cite{phonon} we set 
$|t_1|$=0.2 eV.
 
%\bibitem{plee}
%P.~A. Lee and X.-G. Wen, Phys. Rev. B {\bf 78}, 144517 (2008).


%\bibitem{two} 
%M.A.N.~Ara\'ujo,{\it et al.}, 
%New J. Phys. {\bf 11}, 113008 (2009);
% M.A.N.~Ara\'ujo, P.D.~Sacramento, Phys. Rev. B.{\bf 79}, 174529 (2009);
%M.J.~Calder\'on,{\it et al.}, New J. of Phys. {\bf 11}, 013051 (2009);
% W.-Q.~Chen, {\it et al.}, Phys .Rev. Lett., {\bf 102},047006 (2009);
%A.~Ciechan, K.I.~Wysoki\'nski, Phys. Rev. B., {\bf 80}, 224523 (2009);
%%\bibitem{Daghofer} M. Daghofer, A. Moreo, J.A.Riera, E.Arrigoni,
%%D.J.Scalapino, E.Dagotto, Phys. Rev. Lett. {\bf 101}, 237004 (2008).
%Y.~Gao, W.-P.~Zhu, Phys.Rev.B. {\bf 81}, 104504 (2010);
%P.~Ghaemi, {\it et al.}, Phys. Rev. Lett., {\bf 102}, 157002 (2009);
%X.Hu,{\it et al.},  Phys. Rev. B,{\bf 80}, 014523 (2009);
%H.~Jiang, {\it et al.}, Phys. Rev. B. {\bf 80}, 134505 (2009);
%K.~Kubo, P.~Thalmeier, J. Phy. Soc. Jap, {\bf 78}, 083704 (2009);
%J.~Lorenzana, {\it et al.}, Phys. Rev. Lett. {\bf 101}, 186402 (2008);
%F.~Lu, L.-J.~Zou, J. Phys. {\bf 25}, 255701 (2009);
% T.A.~Maier, D.J.~Scalapino, Phys. Rev. B. {\bf 78}, 020514(R) (2008);
%%\bibitem{Moreo} A.Moreo, M.Daghofer,A.Nicholson,E.Dagotto,
%%Phys.Rev.B {\bf 80}, 104507 (2009)
%%\bibitem{Moreo2} A.Moreo,M.Daghofer,J.A.Riera,E.Dagotto,
%%Phys.Rev.B. {\bf 79}, 134502 (2009)
%M.M.~Parish, {\it et al.}, Phys. Rev.B. {\bf 78}, 144514 (2008);
%E.~Plamadeala, {\it et al.}, Phys. Rev. B {\bf 81}, 134513 (2010);
%P.~Prelovsek, {\it et al.}, Phys. Rev. B.{\bf 80}, 014517 (2009);
%P.~Prelovsek, I.~Sega, Phys. Rev. B. {\bf 81}, 115121 (2010);
%Y.~Ran, {\it et al.}, Phys. Rev. B, {\bf 79} 014505 (2009);
%J.J.~Rodr\'iguez-N\'unez, {\it et al.}, 
%J. Supercond. Nov. Magn. {\bf22}, 539 (2009);
%K.~Seo, {\it et al.},  Phys. Rev. Lett., {\bf 101}, 206404 (2008);
%K.~Seo, {\it et al.}, Phys. Rev. B,{\bf 79}, 235207 (2009);
%R.~Sknepnek, {\it et al.}, Phys. Rev. B,{\bf 79}, 054511 (2009);
%%\bibitem{stanescu} T.D.~Stanescu, V.~Galizski, S.D.~Sarma, Phys. Rev.B.,
%%{\bf 78}, 195114 (2008).
%W.-F.~Tsai, {\it et al.}, Phys. Rev. B,{\bf 80}, 064513 (2009);
%Y.~Wan, Q.-H.~Wang, Euro. Phys. Lett.,{\bf 85}, 57007(2009);
%J.C.~Xavier, {\it et al.}, Phys. Rev. B, {\bf 81}, 085106  (2010);
%Z.J.~Yao, {\it et al.}, New J. of Phys. {\bf 11}, 025009 (2009);
%D.~Zhang, Phys. Rev. Lett, {\bf 103}, 186402 (2009);
%Y.-Y.~Zhang, {\it et al.}, Phys. Rev. B,{\bf 80}, 094528 (2009);
%Y.~Zhou, {\it et al.},  Phys. Rev. B, {\bf 78},
%064514 (2008).
%%\bibitem{RMP01}
%%E. Dagotto, T. Hotta, and A. Moreo, Phys. Rep. {\bf 344},  1   (2001).

\bibitem{oles} 
%A.M. Ole\'s, L.F. Feiner, and J. Zaanen,
%Phys. Rev. B \textbf{61}, 6257 (2000);
A.M. Ole\'s {\it et al.}, {\it ibid.}
{\bf 72}, 214431 (2005).

\bibitem{suple} Results provided in the Supplementary Information material
located at {\it http://sces.phys.utk.edu/bsup}.


\bibitem{foot1} For $J_{\rm H}/U$=$0.3$ $(0.1)$ the phase diagram Fig.~\ref{Fig2}(b) 
was found to be
qualitatively similar to the  $J_{\rm H}/U$=$0.2$.
% case except that the 
%``No Binding'' region shrinks (expands).


\bibitem{rong} R. Yu {\it et al.}, Phys. Rev. B {\bf 79},  104510 (2009); Q. Luo {\it et al.}, Phys. Rev. B {\bf 82}, 104508 (2010).
%arXiv:1007.1436.

\bibitem{teo}
I. Mazin {\it et al.},  Phys. Rev. Lett. {\bf 101},  057003 (2008); \\
K. Kuroki {\it et al.},   Phys. Rev. Lett. {\bf 101},  087004  (2008).
%M.~M. Korshunov and I. Eremin, Phys. Rev. B {\bf 78},  140509  (2008).
%D. Parker, {\it et al.},  Phys. Rev. B {\bf 78},  134524  (2008).%

\bibitem{our3} M. Daghofer {\it et al.}, 
%A. Moreo, A. Nicholson, and E. Dagotto, 
Phys. Rev. B {\bf 81}, 014511 (2010).

\bibitem{nature} B. Zeng {\it et al.}, Nature Comm. {\bf 1}, 112 (2010).


\bibitem{lda+dmft} 
M. Aichhorn {\it et al.}, Phys. Rev. B {\bf 80}, 085101 (2009);
P. Hansmann {\it et al.}, Phys. Rev. Lett. {\bf 104}, 197002 (2010).

%\bibitem{foot2}
%For very large (and unphysical) values of $|E_{\rm B}|$,
%the $B_{\rm 1g}$ symmetry becomes stable for all values of 
%$U$. In this case an operator of the form 
%$(\cos k_x\cos k_y)\sigma_3$ prevails. 
%%This pairing operator is different from the
%%$B_{\rm 1g}$ that characterizes the cuprates that 
%%in this context would be
%%{\rm (ix)} 
%%$(\cos k_x-\cos k_y)\sigma_0$, which has a finite, but much smaller, 
%%overlap than {\rm (viii)}.



\bibitem{jose} E. Dagotto {\it et al.},
%J. Riera, and A.P. Young, 
Phys. Rev. B {\bf 42}, 2347 (1990).

\bibitem{foot} $P(\omega)$'s for $J_{\rm NN}/U$=$0.20$,
inside the $A_{\rm 1g}$ pairing region, were also studied. No 
qualitative changes were found.
% since the low-lying levels cross with each other but 
%remain quasi-degenerate in the studied range of $J_{\rm NN}$.
%Additional electron doping, i.e., $N$=$20$ was
%also studied and
%a similar situation was observed: the $N=20$ GS also has 
%low-lying excited states with many symmetries.

%\bibitem{otherfoot} The equal-time 
%pair-pair correlations at various distances
%lead to conclusions similar to those from $P(\omega)$. 

%\bibitem{newcomment} 
%Increasing $J_{\rm NN}$ at fixed intermediate $U$
%or increasing $U$ at $J_{\rm NN}$=0 are {\it not} equivalent: the former has
%several competing channels while the latter is dominated by $A_{\rm 1g}$.


%\bibitem{commentRPA} We also performed RPA calculations for the two-orbital Hubbard model,
%following Ref.~\cite{graser}, and obtained results in qualitative agreement with the small
%clusters studied. Details will be presented elsewhere.

\end{thebibliography}
\end{document}